\documentclass[aps,prl, twocolumn,superscriptaddress,letterpaper]{revtex4}

\usepackage{graphicx}
\usepackage{amssymb,amsmath}
\usepackage{chemarrow}
\usepackage{chemarr}
\usepackage{float}
\usepackage{bm}
\usepackage{setspace}
\usepackage{extarrows}
\usepackage{parskip}
\usepackage{setspace}
\usepackage{parskip}
\usepackage{xcolor}
\definecolor{midnightblue}{cmyk}{1,1,0,0.1}
\definecolor{forestgreen}{cmyk}{0.76,0,0.26,0.5}

\usepackage{epsfig}
\usepackage{lineno}
\makeatother

\usepackage{hyperref}
\hypersetup{
    bookmarks=true,         
    unicode=false,          
    pdftoolbar=true,        
    pdfmenubar=true,        
    pdffitwindow=false,     
    pdfstartview={FitH},    
    pdftitle={My title},    
    pdfauthor={Author},     
    pdfsubject={Subject},   
    pdfcreator={Creator},   
    pdfproducer={Producer}, 
    pdfkeywords={keyword1} {key2} {key3}, 
    pdfnewwindow=true,      
    colorlinks=true,       
    linkcolor=midnightblue,          
    citecolor=magenta,        
    filecolor=midnightblue,      
    urlcolor=midnightblue,          
}

\begin{document}

\title{Chiral Phonon Activated Spin Seebeck Effect}

\author{Xiao Li}
\affiliation{NNU-SULI Thermal Energy Research Center (NSTER) and Center for
  Quantum Transport and Thermal Energy Science (CQTES), School of Physics and Technology, Nanjing Normal
  University, Nanjing 210023, China}

\author{Jinxin Zhong}
\affiliation{Center for Phononics and Thermal Energy Science, China-EU Joint Lab for Nanophononics,School of Physics Science and Engineering, Tongji University, Shanghai 200092, China}

\author{Jinluo Cheng}
\affiliation{Changchun Institute of Optics, Fine Mechanics and
  Physics, Chinese Academy of Science, Changchun 130033, China}

\author{Hao Chen}
\affiliation{NNU-SULI Thermal Energy Research Center (NSTER) and Center for
  Quantum Transport and Thermal Energy Science (CQTES), School of Physics and Technology, Nanjing Normal
  University, Nanjing 210023, China}
  
\author{Lifa Zhang}
\email{phyzlf@njnu.edu.cn}
\affiliation{NNU-SULI Thermal Energy Research Center (NSTER) and Center for
  Quantum Transport and Thermal Energy Science (CQTES), School of Physics and Technology, Nanjing Normal
  University, Nanjing 210023, China}

\author{Jun Zhou}
\email{zhoujunzhou@njnu.edu.cn}
\affiliation{NNU-SULI Thermal Energy Research Center (NSTER) and Center for
  Quantum Transport and Thermal Energy Science (CQTES), School of Physics and Technology, Nanjing Normal
  University, Nanjing 210023, China}


\begin{abstract}
Efficient generation of spin polarization is the central focus of spintronics. In magnetic materials, spin currents can arise from heat currents by the conventional spin Seebeck effect. Recently, chiral phonons with definite handedness and angular momenta have also produced profound impacts on multiple research fields. In this paper, starting with nonequilibrium distribution of chiral phonons under temperature gradient, we find a new spin selectivity effect - chiral phonon activated spin Seebeck (CPASS) effect, in chiral materials without magnetic order nor spin-orbit coupling. With both phonon-drag and band transport contributions, the CPASS coefficients are computed based on the Boltzmann transport theory. The spin accumulations by the CPASS effect quadratically increase with temperature gradient, and vary with the chemical potential modulation, thus enabling highly efficient and tunable spin generation. The CPASS effect provides a promising explanation on the chiral-induced spin selectivity effect and opportunities for designing advanced spintronic devices based on nonmagnetic chiral materials.   
\end{abstract}

\maketitle

\textcolor{forestgreen}{\emph{\textsf{Introduction}.}}---
The magnetism induced by phonons in solids has been ignored for a long time since phonons were traditionally regarded as linearly polarized and did not carry angular momentum. This circumstance has been changed since the discovery of nonzero angular momentum in chiral materials 
\cite{zhang2014,zhang2015,juraschek2019,nova2017,zhu2018}. The ``vibrational angular momentum'' of chiral phonons, which describes 
the coupled-vibration-rotation of ions, was found to be important in nonequilibrium state when a temperature gradient is applied as an example \cite{hamada2018}.
The net magnetic moment induced by chiral phonons was firstly thought to be much smaller than the Bohr magneton 
\cite{hamada2018,juraschek2019}. 
Lately, unexpected large magnetic moment was found in experiment \cite{cheng2020,kim2021}. 
It was possibly attributed to the phonon-modified electronic energy together with the 
momentum-space Berry curvature \cite{ren2021}. 

We recognize that such large magnetic moment can be used to generate 
spin-polarized electrons which plays a crucial role in the field of
spintronics \cite{spintronicsbook}. Usually, the magnetism and/or spin-orbit interaction are indispensable
for the manipulation the electron spin. People found that the chirality of materials, 
which characterizes the mirror asymmetry, provides an alternative source of the spin polarization \cite{rikken2011}. 
As early as 1995, Mayer and Kessler \cite{mayer1995} measured the spin-dependent attenuation of
electron beams through a vapor of chiral molecules due to the
spin-orbit coupling. However, the measured signal was found to be
extremely low. In 2011, G\"{o}hler \cite{gohler2011} reported a much
stronger spin filter effect when electrons transport through
double-stranded DNA. After that, the widely observed spin polarization in
various chiral materials led to the rise of a field called the
chiral-induced spin selectivity (CISS)
effect \cite{naaman2015,michaeli2017}. The mechanism of the CISS effect is still
not very clear because there is no magnetic field and the spin-orbit coupling
in organic molecules is very weak due to the absence of heavy elements.
Different origins of the CISS effect were proposed, e.g. the electric
field pumping effect \cite{rai2013}, the electron transmission
through helical potential \cite{eremko2013,gutierrez2012}, the
interplay of the spin-orbit coupling and dipolar potential
\cite{michaeli2019}. We believe that the CISS effect could possibly be induced by chiral phonons because the 
electron spin could be manipulated by the magnetic moment generated by the phonon angular momentum, i.e. coupled to the phonon angular momentum in a similar way to the spin-rotation coupling
\cite{matsuo2013,hamada2015}.

It is well known that the temperature gradient is able to generate spin
voltage in magnets and materials with the sizable spin-orbit coupling. Such effect is called the spin Seebeck
effect \cite{uchida2008,adachi2013}. Is it possible to utilize the
temperature gradient to generate spin voltage in chiral materials in
the absence of the magnetic order and spin-orbit coupling?

\begin{figure}[h!]
	\centering
	\includegraphics[width=6.5 cm]{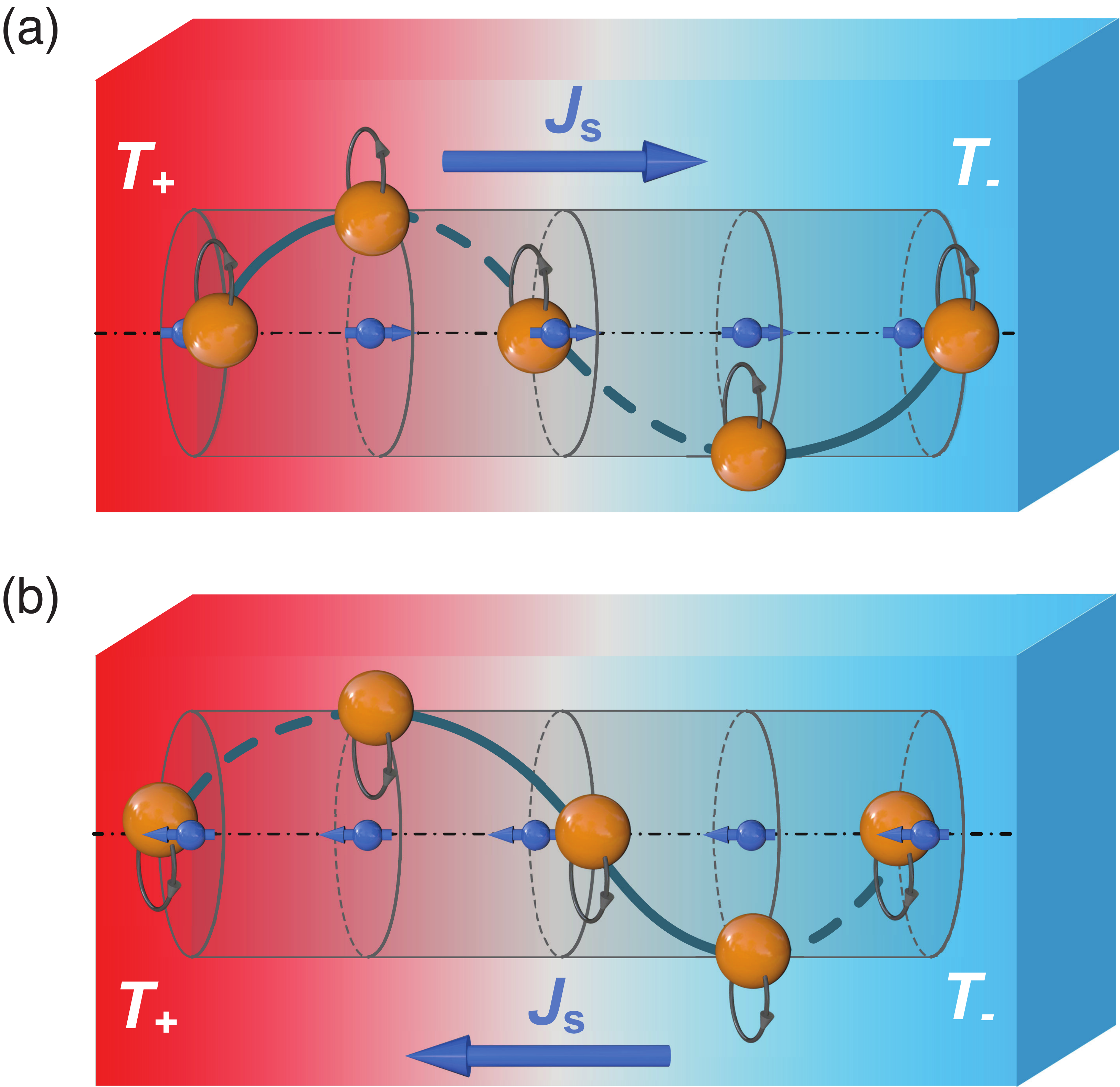}
	\caption{Schematic depiction of the CPASS effect in chiral materials. (a) and (b) correspond to chiral phonons with opposite circular motions of nuclei, giving rise to opposite spin polarization of electrons, respectively. The blue and yellow balls stand for electrons and nuclei in the chiral material, respectively. The grey arcs represent atomic circular motions with arrows denoting the directions of the motions. The arrows on electrons point to the spin directions. A red-to-blue gradient background stands for an applied temperature gradient.}
	\label{Sche}
\end{figure}

In this paper, we theoretically investigate the chiral phonon activated spin Seebeck (CPASS) effect due to
nonequilibrium distribution of chiral phonons under a temperature gradient in
chiral materials, by the Boltzmann transport theory. The temperature gradient not only lifts the spin
degeneracy in the band structure but also plays as the driving force of spin accumulation or spin current as shown in Fig. \ref{Sche}. 
As a result, a net spin accumulation is created, with both the phonon-drag and band transport contributions. 
The computed CPASS coefficients linearly depend on the temperature gradient and correspondingly the 
spin accumulation is found to nearly quadratically increase with the temperature gradient.
The phonon-drag and band transport contributions are opposite with tunable relative magnitude by shifting the chemical potential. 
This new mechanism of the spin Seebeck effect proposed here gives full play to chiral phonons in spin-polarized transport, and provides another possible origin of the CISS effect even in the absence of any magnetic order and strong spin-orbit coupling.


\textcolor{forestgreen}{\emph{\textsf{Theoretical model}.}}--- 
Due to the lack of inversion, mirror symmetry, rotoreflection symmetry, the phonons in chiral materials present specific chirality. The chirality of phonons and associated phonon angular momentum are locked with the chirality of materials, and the reversal of the material chirality is accompanied by the reversals of the phonon chirality and phonon angular momentum \cite{Chen2021}. On the other hand, it was also found that a net phonon angular momentum can be produced in chiral materials by non-equilibrium distribution of chiral phonons under a temperature gradient and the angular momentum exhibits a linear dependence on the temperature gradient \cite{hamada2018,hamada2020}. The net phonon angular momentum can exert an effective magnetic field on Bloch electrons, which introduces opposite Zeeman-type energy shifts in the electronic band structure for opposite spins. Therefore, the above spin-dependent energy shift in chiral materials linearly depends on the temperature gradient and it can be
written as, $\sigma\kappa\nabla T/2$. 
Here, the spin index, $\sigma=\pm 1$, denotes opposite spins and $\nabla T$ is the temperature gradient. $\kappa$ is a material-specific strengths coefficient of the spin splitting. Its sign is determined by the chirality of the material and its magnitude can be estimated from experimental measurement and model calculation, which will be discussed in the next section.

The spin splitting term explicitly exhibits its dependence on the temperature gradient and the material chirality. It is also noted that the spin splitting here is not contributed from the spin-orbit coupling and applies to a chiral material composed of light elements as well. 
Further considering an isotropic parabolic dispersion, the electronic band energy reads,
\begin{equation}
	\epsilon_\sigma(\textbf k)=\hbar^2k^2/2m+\sigma\kappa\nabla T/2,
	\label{band}
\end{equation}
where $m$ is the electron effective mass, and  $\textbf k$ denote the wave vector of Bloch electron, which, together with the subscript $\sigma$, labels the electronic state.

Given the above spin splitting in the electronic structure, we investigate spin-dependent transport behavior under the temperature gradient using the Boltzmann transport equations \cite{Mahan2014}. For the collision processes, the electron-phonon scattering is explicitly considered, which leads to a phonon-drag current. Further taking into account the open circuit without net electron current for both spins, the chiral phonon activated spin accumulation generated in non-magnetic chiral materials can be derived. The methods of the formula derivation and the subsequent numerical calculation are given in the section $\textit{Methods}$.

For the electronic band dispersion in Eq. (\ref{band}), the derived spin accumulation arising from the CPASS effect is longnitudinal and isotropic, and it has a form as
\begin{equation}
	\nabla \delta \mu \equiv \nabla  (\mu_{+}-\mu_{-})=-\left(\frac{L^{pd}_+}{L_+}\!-\!\frac{L^{pd}_-}{L_-}\right)\nabla T-\left(\frac{L^b_+}{L_+}\!-\!\frac{L^b_-}{L_-}\right)\nabla T
	\label{saccu}
\end{equation}
where $\mu_\sigma$ is the chemical potential with $\sigma$ distinguishing opposite spins and $T$ is the temperature. $\nabla \mu_\sigma$ and $\nabla T$ are corresponding spatial gradients, and they are assumed to be along $x$-axis for simplicity to calculate the longitudinal and isotropic CPASS effect. $L^{pd}_{\sigma}$ is an integral associated with the phonon-drag effect arising from the electron-phonon scattering, while $L^{b}_{\sigma}$ explicitly includes the electronic band energy but excludes phonon-related terms. $L_{\sigma}$ is a longitudinal conductivity of a unit charge. Therefore, two terms of the right hand side of Eq. (\ref{saccu}) correspond to the contributions from the phonon-drag effect and the electronic band transport, respectively.

The integrals, $L^{pd}_{\sigma}$, $L^{b}_{\sigma}$ and $L_{\sigma}$, are written as,
\begin{linenomath}\begin{subequations}
	\begin{align}
		L^{pd}_{\sigma}&=-\frac{1}{\nabla T}\int\frac{d^{3}\textbf k}{(2\pi)^{3}} v_{\sigma,x}(\textbf {k})\tau_\sigma(\textbf {k})\frac{\partial f_{\sigma}(\textbf k)}{\partial t}\bigg|_{pd} \label{L1} \\
		L^b_\sigma&=\frac{1}{T}\int\frac{d^{3}\textbf k}{(2\pi)^{3}}v^2_{\sigma,x}(\textbf {k})\tau_\sigma(\textbf {k})[\epsilon_\sigma(\textbf k)-\mu_\sigma][-\frac{\partial f^0_{\sigma}(\textbf {k})}{\partial \epsilon_\sigma(\textbf k)}] \label{L2} \\
		L_{\sigma}&=\int\frac{d^{3}\textbf k}{(2\pi)^{3}}v^2_{\sigma,x}(\textbf {k})\tau_\sigma(\textbf {k})[-\frac{\partial f^0_{\sigma}(\textbf {k})}{\partial \epsilon_\sigma(\textbf k)}] \label{L3}
	\end{align}
	\label{L123}
\end{subequations}\end{linenomath}
where $v_{\sigma,x}$ is the $x$ component of the electron's group velocity. $\tau_{\sigma}(\textbf {k})$ is a total electron relaxation time, including contributions from multiple scattering processes, e.g. the electron-phonon scattering and the impurity scattering. $f_{\sigma}(\textbf k)$ and $f^{0}_{\sigma}(\textbf k)$ are the nonequilibrium distribution function and the equilibrium Fermi-Dirac distribution function for electrons, respectively. $f^{0}_{\sigma}(\textbf k)=[e^{\frac{\epsilon_\sigma(\textbf k)-\mu_\sigma}{k_BT}}+1]^{-1}$, with $k_B$ being the Boltzmann constant. $\frac{\partial f_{\sigma}(\textbf k)}{\partial t}|_{pd}$ is the collision integral from the phonon-drag effect, and its expression is given in the section $\textit{Methods}$. Based on Eqs. (\ref{saccu}) and (\ref{L123}), the total CPASS coefficient can be further obtained as, $S_{tot}=\nabla \delta \mu/\nabla T$, with two corresponding contributions denoted by $S_{pd}$ and $S_{b}$.



For the above CPASS effect, it is noted that phonon modes are involved in two processes: 1) inducing spin splitting in the band structure; 2) generating phonon-drag current.
The phonon modes involved in the two processes are not necessary to be the same.
For the phonon-drag current,
the longitudinal acoustic phonon is regarded as a major contributor,
since the longitudinal acoustic phonon has a larger non-equilibrium  population compared with optical branches, associated with its larger equilibrium distribution and larger group velocity, according to the phonon Boltzmann transport equation (see Eq. \ref{SBE2} of the section $\textit{Methods}$). In contrast, chiral phonons with atomic circular motions, inducing nonzero phonon angular momentum and associated spin splitting, are considered to exist in the entire phonon Brillouin zone except for the high-symmetry momentum points, e.g. the Brillouin zone center with the time-reversal symmetry \cite{hamada2018}. Each phonon mode's contribution to the spin splitting depends on not only its non-equilibrium population but also the magnitude of its associated phonon angular momentum. Therefore, we take into account only longitudinal acoustic modes in the calculation of the phonon-drag current for simplicity, while the contributions from all chiral phonons are embodied in the parameter $\kappa$ that determines the magnitude of the spin splitting in the electronic band structure. By introducing the scattering of electrons by the longitudinal acoustic phonons into the term $\frac{\partial f_{\sigma}(\textbf k)}{\partial t}|_{pd}$ in Eq. (\ref{L1}), the numerical calculations of the CPASS effect are performed. The calculation results are demonstrated and discussed below.

\begin{figure}[ht]
	\centering
	\includegraphics[width=8.0 cm]{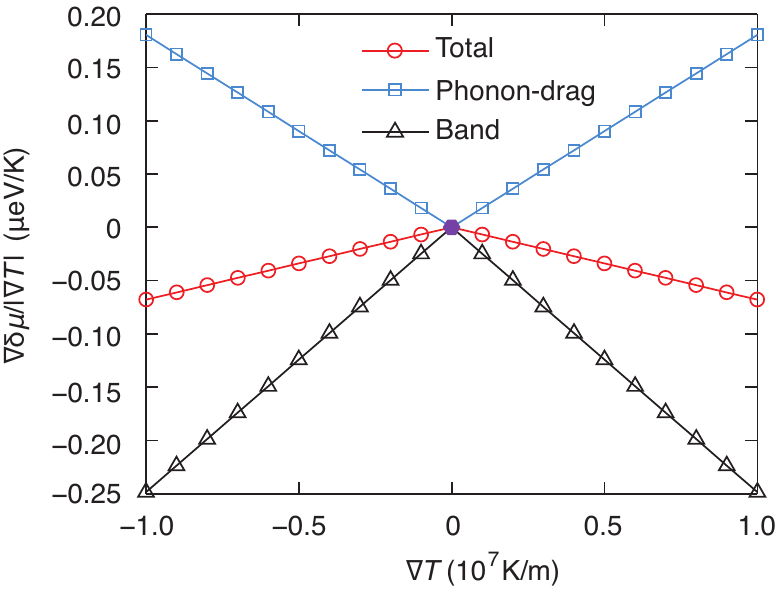}\caption{Changes of the CPASS coefficients with the temperature gradient. The phonon-drag and band transport contributions are denoted by blue square and black triangle, respectively, while the total contribution is given by red circle. The purple hexagon represents the point that is located on linear curves of the CPASS coefficients but numerically meaningless due to the division by zero. The parameters, $\mu=0.2$ eV,  $\kappa=10^{-10}$ eV$\cdot$m/K and $T=300$ K.}
	\label{nablaT}
\end{figure}

\textcolor{forestgreen}{\emph{\textsf{Numerical results}.}}---
Figure \ref{nablaT} shows representative evolutions of the CPASS coefficients with the temperature gradient $\nabla T$, where the band transport contribution, the phonon-drag contribution and their sum, i.e. the total contribution are all plotted. Without loss of generality, we set $\kappa=10^{-10}$ eV$\cdot$m/K, which means a Zeeman splitting of the order of 0.1 meV can be induced in electronic band structure when a temperature difference of 1 K is applied to a sample of 1 micro meter thickness, where an effective magnetic field of 1 T is generated. The detail of the evaluation can be found in Supplemental Material. The considerable effective magnetic field arising from chiral phonons agrees with the measurement of the effective phonon magnetic moment \cite{cheng2020}, which validates the choice of the $\kappa$. While the typical value of the $\kappa$ is used here, the evolution of the CPASS coefficients with $\kappa$ will be discussed later. 

Using reasonable material-specific parameters, the computed CPASS coefficients in Figure \ref{nablaT} are considerable, and they are expected to be measurable by readily available experimental techniques such as magneto-optic Kerr effect. It is also found that the magnitudes of the CPASS coefficients increase linearly with the increase of the temperature gradient. That is, the spin accumulation, $\nabla\delta \mu$, has a quadratic dependency on the temperature gradient. This is because besides the explicit $\nabla T$ in Eq. (\ref{saccu}),  $\nabla T$ is also included in $\epsilon_\sigma(\textbf k)$ with the spin splitting term arising from the nonequilibrium phonon distribution under the spatial temperature difference. In contrast, if we assume that the spin splitting is independent of the temperature gradient, the calculated CPASS coefficients will be constant. The CPASS effect is thus more sensitive to the temperature gradient, enabling the efficient generation of the spin polarization under the action of the temperature gradient.    
Besides, the band transport contribution and the phonon-drag contribution are always opposite in the parameter space considered when the carriers' charge is regarded as negative, and their relative magnitudes vary with the parameter choices. Moreover, when the temperature gradient is reversed, i.e. $\nabla T\rightarrow -\nabla T$, the calculated $\delta \mu$ and $\delta \mu/|\nabla T|$ are unchanged as shown in Fig. \ref{nablaT}, since the right hand side of Eq. (\ref{saccu}) does not change if we let $\sigma \rightarrow -\sigma$.

\begin{figure}[ht]
	\centering
	\includegraphics[width=8.0 cm]{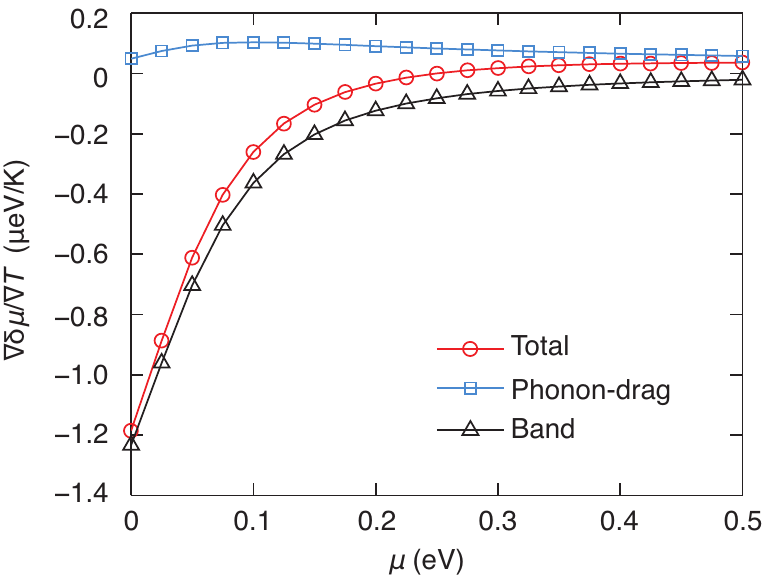}\caption{Changes of the CPASS coefficients with the varied chemical potential. The parameters, $\nabla T=5\times10^{6}$ K/m,  $\kappa=10^{-10}$ eV$\cdot$m/K and $T=300$ K. }
	\label{mu}
\end{figure}

Figure \ref{mu} demonstrates the changes of the CPASS coefficients at a given temperature gradient with the chemical potential, $\mu=(\mu_{+}+\mu_{-})/2$. When the chemical potential is chosen be close to the band edge ($\mu\approx 0$ eV), i.e. the system is in the non-degenerate limit, it is found that the band transport contribution to the CPASS coefficient is much larger than the phonon-drag contribution by tens of times.  As the chemical potential is shifted upwards, the band transport contribution always decreases, by up to more than one order of magnitude, while the phonon-drag contribution varies within the same order.  As a result, the band transport contribution becomes closer to and then smaller than the phonon-drag contribution in magnitude. Given that the two contributions are opposite, a switch of their relative magnitudes leads to a sign change of the total CPASS coefficient. Therefore, the CPASS coefficients are highly tunable with the modulation of the chemical potential. Moreover, it is seen that the $\mu$-sensitive variation in the band transport contribution arises from the difference between the electronic band energy $\epsilon_\sigma(\textbf k)$ and the chemical potential in Eq. \ref{L2}. When moving the chemical potential away from the band edge, the difference is not always positive for varied electronic energies, leading to the cancellation of the differences with opposite signs and corresponding decreased band transport contribution to the CPASS effect.

\begin{figure}[ht]
	\centering
	\includegraphics[width=8.0 cm]{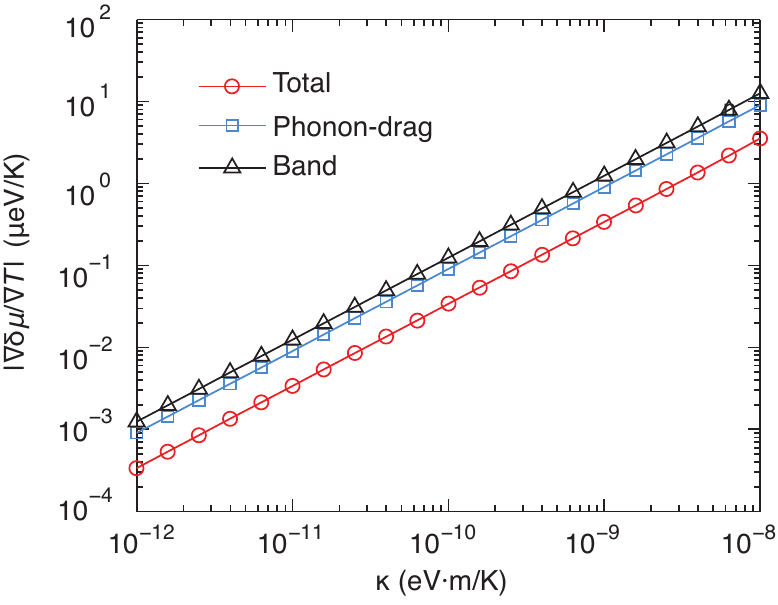}\caption{Changes in magnitudes of the CPASS coefficients with the varied strength coefficient of the spin splitting. Both axes are plotted using a logarithmic scale. The parameters, $\mu=0.2$ eV, $\nabla T=5\times10^{6}$ K/m and $T=300$ K.}
	\label{kappa}
\end{figure}

The magnitudes of the CPASS coefficients are further computed as functions of the strength coefficient of the spin splitting, $\kappa$, that is included in the spin splitting term of Eq. \ref{band}, while the other parameters are kept unchanged. They are demonstrated in Figure \ref{kappa}, by a log-log scaled plot. It is seen that the strengths of the CPASS effects, including the phonon-drag and band transport contributions, are all linearly proportional to $\kappa$ and associated spin splitting in the band structure, over a large range of $\kappa$, according to the lines with the slope of 1 in the log-log plot. When $\kappa=0$, i.e. the electronic band structure has no spin splitting, the computed CPASS coefficients are vanishing as well. Therefore, the spin splitting induced by the nonequlibrium distribution of chiral phonons under the temperature gradient is necessary for the CPASS effect in chiral materials. Besides, it is noted when the sign of $\kappa$ is changed, the calculated spin accumulation reverses because the last term in Eq. (\ref{band}) is unchanged when $\sigma \rightarrow -\sigma$ and $\kappa \rightarrow -\kappa$. The left-handed and right-handed materials thus lead to opposite spin polarizations under the same temperature gradient, as illustrated in Fig. \ref{Sche}.

Given that the phonon angular momentum generated under the action of the temperature gradient is inversely proportional to the square of the temperature \cite{hamada2018}, the strength coefficient, $\kappa$, is considered to exhibit the same dependency on the temperature with the phonon angular momentum. Besides the $\kappa$, the temperature dependency also exists in other parts of the above formulas of the CPASS coefficients, e.g. the distribution functions of electron and phonon. Figure 5 demonstrates the variations of the CPASS coefficients with the temperature. The solid lines consider all temperature dependencies, from both the $\kappa$ and the other parts, while the dashed lines consider a temperature-independent $\kappa$. By comparing the two cases, the importance of the temperature dependency of the $\kappa$ will become clear. When the temperature-independent $\kappa$ is taken into account, the band transport contribution to the CPASS coefficient is enhanced in magnitude with the increase of the temperature. In contrast, the phonon-drag contribution is almost unchanged with the temperature. As a result, the total CPASS coefficient exhibits the same trend with that of the band transport contribution.
After the temperature dependency of the $\kappa$ is also included, both the band transport contribution and the phonon-drag contribution decrease as the temperature increases, due to the decreased $\kappa$ and associated smaller spin splitting. However, the difference in magnitude between two contributions increases, leading to an enhanced magnitude of the total CPASS coefficient.

\begin{figure}[ht]
	\centering
	\includegraphics[width=8.0 cm]{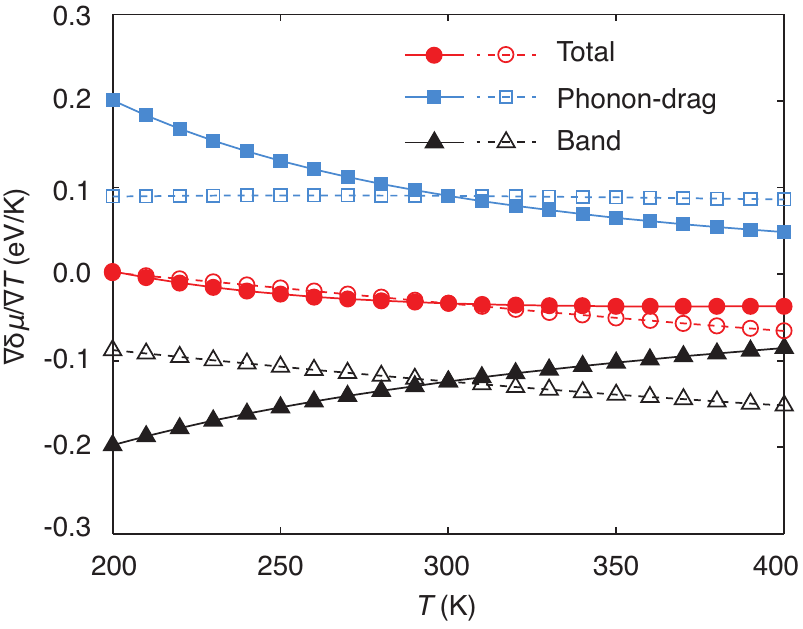}\caption{Changes of CPASS coefficients with increasing temperature. The solid and dashed lines correspond to the temperature-dependent and temperature-independent $\kappa$, respectively. The parameters, $\mu=0.2$ eV and $\nabla T=5\times10^{6}$ K/m. Both the temperature-independent $\kappa$ and temperature-dependent $\kappa$ at 300 K adopt the value of $\kappa=10^{-10}$ eV$\cdot$m/K.}
	\label{T}
\end{figure}

\textcolor{forestgreen}{\emph{\textsf{Discussion}.}}---
While the above numerical calculations exhibit the dependencies of the CPASS effect on the temperature gradient, the chemical potential, the material-specific spin splitting and the temperature, the analytic expressions for the CPASS coefficients are also derived in degenerate and non-degenerate limits and provide further insights into the compelling effect. The detailed derivation is provided in Supplemental Material. In the degenerate limit, i.e. the average chemical potential $\mu$ is much larger than the thermal energy $k_BT$, both the phonon-drag and band transport contributions to the CPASS coefficients are obtained by considering the scattering of electrons by the longitudinal acoustic phonons and applying the Sommerfeld expansion to integrals in Eqs. (\ref{L1})-(\ref{L3}). They are expressed as,
\begin{linenomath}\begin{subequations}
	\begin{align}
		S_{pd}&=-\frac{D^2_ac^2_{L}\tau_L m^{\frac{5}{2}}(\textbf e_q\cdot\textbf q)^2}{2\pi\rho\hbar^2 k_BT^2\sqrt{2\mu}} \frac{\partial F(k)}{\partial k}\bigg|_{k_F}\kappa\nabla T \label{pdl1} \\
		S_b&=\frac{\pi^2k_B^2T}{2 u^2}{\delta u} \label{bl1}
	\end{align}
\end{subequations}\end{linenomath}
where $D_a$, $\textbf e_q$ and  $\rho$ being the deformation potential, the unit vector pointing to the direction of the phonon displacement and the mass density of the chiral material considered, respectively. $c_{L}$ and $\tau_L$ are the momentum-independent group velocity and the average relaxation time of the longitudinal acoustic phonon, respectively. $F(k)$ is a function of the magnitude of the wave vector $\textbf k$ (see Supplemental Material) and $k_F$ is the magnitude of the Fermi wave vector. Considering $k_BT\ll\mu$ and $\delta\mu\ll\mu$ in the degenerate limit, the band transport contribution, $S_b$, is small. As the dominating contribution, $S_{pd}$ exhibits an explicit linear dependency on $\nabla T$ and $\kappa$, which, together with the relative magnitude of two contributions in the limit, agrees with the above numerical calculations. Besides, it is seen that $S_{pd}$ depends on more parameters associated with the Bloch electron and the longitudinal acoustic phonon, demonstrating distinct scaling laws. The enhancement of $S_{pd}$ prefers the chiral materials with large deformation potential, large group velocity and average relaxation time of the longitudinal acoustic phonon, and large effective mass of the Bloch electron, but with small mass density. Moreover, since $\frac{\partial F(k)}{\partial k}\big|_{k_F}$ in Eq. \ref{pdl1} is likely to include $k_F$ that is also related with the temperature gradient by the energy dispersion relationship in Eq. \ref{band}, higher-order terms in the temperature gradient probably exists in the spin Seebeck coefficient.

In the non-degenerate limit, by replacing the Fermi-Dirac distribution of electrons with the Boltzmann distribution and considering a small spin splitting induced by the temperature gradient, the CPASS coefficient arising from the band transport contribution can be obtained as,
\begin{equation}
	\begin{split}
		S_b=-\frac{\kappa\nabla T}{T}+\frac{\delta\mu}{T} \\
	\end{split}
\end{equation}
It is seen that the linear term in $\kappa$ and $\nabla T$ is also embodied in $S_b$, besides a term that is independent of these variables and related to the chemical potential difference between opposite spins. Considering $\delta\mu$ is small in a nanoscale sample, the above linear term in $\nabla T$ is dominating. The above analytic results in the limit are also consistent with our numerical calculations.

Furthermore, it is noted the CPASS effect proposed in chiral materials is a new mechanism that is completely different from the conventional spin Seebeck effect. From the aspect of the physical origin, the CPASS effect is rooted in chiral phonons under the temperature gradient, and it does not require the magnetic order nor the spin-orbit coupling. On the other hand, as a consequence of the physical origin, the CPASS effect enables a highly efficient spin generation with a quadratic dependency of the temperature gradient, compared with the conventional spin Seebeck effect. The new characteristics provides a promising explanation on the CISS  observed in chiral materials where both the magnetic order and strong spin-orbit coupling are absent, and provides opportunities for the exploration of advanced spintronic devices based on chiral materials. The CPASS effect can be observed by inverse spin-Hall effect or transient magneto-optic Kerr measurements on chiral semiconductors with low thermal conductivity subject to a temperature gradient \cite{kim2021}. Moreover, the spin Seebeck effect proposed here is not limited to chiral materials and it is expected to have generalizations to other non-centrosymmetric materials, including polar materials, since these materials can also obtain net phonon angular momentum and effective magnetic field under a temperature gradient \cite{hamada2018}. While our high-symmetry model has only nonvanishing longitudinal, isotropic CPASS coefficients, distinct CPASS response tensors, including longitudinal and transverse components, are expected in various non-centrosymmetric materials with different crystalline symmetries. It is worth quantitative calculations on the response tensors of more realistic materials by e.g. first-principles methods to provide more material candidates for the CPASS effect.

 \medskip

\textcolor{forestgreen}{\emph{\textsf{Methods}.}}---
The CPASS effect is studied by the Boltzmann transport equations for electrons and phonons \cite{Mahan2014}, which are given as,
\begin{equation}
	\begin{split}
		{\textbf{v}_\sigma}(\textbf k)\cdot & \left\{\nabla\mu_\sigma+[\epsilon_\sigma (\textbf k)-\mu_\sigma]\frac{\nabla T}{T}\right\}\left[-\frac{\partial f^0_\sigma{(\textbf k)}}{\partial\epsilon_\sigma (\textbf k)}\right] \\
		&=-\frac{\delta f_{\sigma}(\textbf {k})}{\tau_{\sigma}(\textbf {k})}+\frac{\partial f_{\sigma}(\textbf k)}{\partial t}\bigg|_{pd}
		\label{BTE}
	\end{split}
\end{equation}
and
\begin{equation}
	\frac{\textbf {v}_{\lambda}(\textbf q)\cdot \nabla T}{k_{B}T^{2}}\hbar\omega_{\lambda}(\textbf{q})n^0_{\lambda}(\textbf q)[n^0_{\lambda}(\textbf q)+1]=-\frac{\delta n_{\lambda}(\textbf q)}{\tau_{\lambda}(\textbf q)}
	\label{SBE2}
\end{equation}
The left hand side and right hand side of the equations describe the drift and collision processes, respectively. In the collision terms, the relaxation-time approximation is adopted for both electrons and phonons, and the phonon-drag effect is also included by the term, $\frac{\partial f_{\sigma}(\textbf k)}{\partial t}|_{pd}$, for electrons. In the above equations, $\textbf{q}$ and $\lambda$ are the wave vector and the band index of phonons, respectively, while $\textbf{k}$ and $\sigma$ are indices of electrons as mentioned above.  
$\textbf{v}_\sigma(\textbf k)$ and $\textbf{v}_\lambda(\textbf q)$ are the group velocities of electron and phonon, respectively.
The electron's distribution function $f_{\sigma}(\textbf {k})$ satisfies the relation, $f_{\sigma}(\textbf {k})=f^{0}_{\sigma}(\textbf k)+\delta f_{\sigma}(\textbf {k})$, where $\delta f_{\sigma}(\textbf {k})$ is the derivation from the equilibrium distribution $f^{0}_{\sigma}(\textbf k)$. Similarly, $\delta n_{\lambda}(\textbf q)$ is the derivation from the phonon equilibrium distribution function $n^{0}_{\lambda}(\textbf q)=[e^{\frac{\hbar\omega_\lambda(\textbf q)}{k_BT}}-1]^{-1}$,
where $\omega_\lambda(\textbf q)$ is the phonon eigenfrequency. $\tau_{\lambda}(\textbf q)$ is the phonon's relaxation time. 

The phonon-drag term, $\frac{\partial f_{\sigma}(\textbf k)}{\partial t}|_{pd}$, is written as, 
\begin{align}
	\frac{\partial f_{\sigma}(\textbf k)}{\partial t}\bigg|_{pd}\!=&\frac{2\pi}{\hbar}\!\sum_{\lambda}\!\int\!\frac{d^{3}\textbf q}{(2\pi)^{3}}|M_{\lambda}(\textbf {q})|^{2}\notag \\ 
	&\times\{[f^0_\sigma(\textbf k\!+\!\textbf q)\!-\! f^0_\sigma(\textbf {k})] \delta(\xi^{\!+\!\textbf q}_{\sigma}) \notag \\ 
	&+[f^0_\sigma(\textbf k\!-\!\textbf q)\!-\! f^0_\sigma(\textbf {k})] \delta(\xi^{\!-\!\textbf q}_{\sigma})\}\delta n_{\lambda}(\textbf q)
	\label{pd}
\end{align}
where $|M_{\lambda}(\textbf {q})|^{2}$ is the matrix element ot the electron-phonon scattering. Taking into account the momentum conservation in the electron-phonon scattering, a phonon with a given $\textbf q$ enables the scatterings between electronic states $\textbf k$ and $\textbf k'= \textbf k\pm \textbf q$. The energy conservation is embodied in two Dirac delta functions, $\delta(\xi^{\pm\textbf q}_{\sigma})$, with the energies, $\xi^{\pm\textbf q}_{\sigma}=\epsilon_{\sigma}(\textbf k)-\epsilon_{\sigma}(\textbf {k}\pm\textbf {q})\pm\hbar\omega_{\lambda}(\textbf {q})$. According to the phonon Boltzmann equation (\ref{SBE2}), $\delta n_{\lambda}(\textbf q)$ in the phonon-drag term is also easily obtained.

An open circuit condition is further considered in the CPASS effect and it reads,
\begin{equation}
	\int\frac{d^{3}\textbf k}{(2\pi)^{3}}\textbf {v}_\sigma(\textbf {k})\delta f_{\sigma}(\textbf {k})=0
\end{equation}
That is, there is no net electron flow.
Considering that $\delta f_{\sigma}(\textbf {k})$ can be expressed by the drift and
the phonon-drag terms according to Eq. (\ref{BTE}) and the CPASS effect is longnitudinal and isotropic for the simple electronic band dispersion in Eq. (\ref{band}), the above condition is rewritten as,
\begin{equation}
	L^{pd}_{\sigma}\nabla T+L^{b}_{\sigma}\nabla T+L_{\sigma} \nabla \mu_\sigma=0
	\label{occ}
\end{equation}
The explicit $\nabla T$ in the first term of the left hand side of the equation is extracted from $\delta n_{\lambda}(\textbf q)$ in $\frac{\partial f_{\sigma}(\textbf k)}{\partial t}|_{pd}$. The forms of the integrals, $L^{pd}_{\sigma}$, $L^{b}_{\sigma}$ and $L_{\sigma}$, are given in the above equations (\ref{L1})-(\ref{L3}). We can then obtain the formulas of the spin accumulation from the CPASS effect, as shown in Eq. (\ref{saccu}), and associated CPASS coefficients. 

In the numerical calculation, for the longitudinal acoustic phonon involved in the electron-phonon scattering, the phonon dispersion is approximately linear  in the neighborhood of the phonon Brillouin zone center,
i.e. $\omega_{L}(\textbf q)=c_{L}q$.
Taking into account the scattering between electrons and longitudinal acoustic phonons, the scattering matrix element reads,
$|M_{L}(\textbf {q})|^2=\frac{D^2_a\hbar (\textbf e_q\cdot\textbf q)^2}{2\rho c_{L} q}$.
By simplifying the delta functions in Eq. (\ref{pd}) that are composed with quadratic terms in $\textbf q/\textbf k$, and setting the Debye wavenumber as the upper limit of the integration with respect to $q$, the spin accumulation generated by the CPASS effect in Eq. (\ref{saccu}) and associated CPASS coefficients are computed numerically. More details of the formula derivation and the numerical calculation can be found in Supplemental Material. The parameter choices are also provided in Supplemental Material, unless otherwise specified in the main text.

\bigskip

\textcolor{forestgreen}{\emph{\textsf{Acknowledgement}.}}---
We acknowledge support from the National Natural Science Foundation of China (No. 11904173, No. 11890703). X.L. is also supported by the Jiangsu Specially-Appointed Professor Program. J.Z. is supported by the Key-Area Research and Development Program of Guangdong Province (No. 2020B010190004).




\end{document}